\documentclass[aps,prb,reprint,groupedaddress,showpacs]{revtex4-1}

\usepackage{graphicx}
\usepackage{dcolumn}

\begin{document}


\title{Thermoelectric studies of K$_{x}$Fe$_{2-y}$Se$_2$: weakly correlated superconductor}


\author{Kefeng Wang}
\author{Hechang Lei}
\author{C. Petrovic}
\affiliation{Condensed Matter Physics and Materials Science Department, Brookhaven National Laboratory, Upton, New York 11973 U.S.A.}

\date{\today}

\begin{abstract}
We report thermal transport properties of K$_x$Fe$_{2-y}$Se$_2$ superconducting single crystal. Peak anomaly in thermal conductivity is observed
at nearly $\frac{T_C}{2}$,  attributed to phonons. The thermoelectric power above $T_c$ exhibits nearly linear behavior and could be described well by the carrier diffusion mechanism in a wide temperature range. The zero-temperature extrapolated thermoelectric power
is smaller than the value in typical strongly correlated superconductors, implying large normalized Fermi temperature. These findings indicate that K$_x$Fe$_{2-y}$Se$_2$ is a weakly or intermediately correlated superconductor.

\end{abstract}
\pacs{74.25.fc, 74.25.fg, 74.20.Mn, 74.70.Xa}

\maketitle

\section{Introduction\label{Introduction}}
The discovery of superconducting transition temperature up to $T_c=26$ K in LaFeAsO$_{1-x}$F$_x$ (1111-type) with $x\sim 0.11$\cite{discovery} generated intense activity. \cite{review1,review2}
Soon it was discovered that superconducting transition temperature can be pushed to higher values by substituting La ions with heavier rare earths, \cite{RE-doped1, RE-doped2, RE-doped3} and the highest $T_c\sim 55$ K were achieved in
SmO$_{1-x}$F$_x$FeAs \cite{highest-Tc1} and $\sim 56.3$ K in Gd$_{1-x}$Th$_x$FeAsO. \cite{highest-Tc2}
Superconductivity was also discovered in the doped AFe$_2$As$_2$ (122-type, A=Ba, Sr, Ca) with ThCr$_2$Si$_2$ structure\cite{122-1,122-2,122-3} that contain a double FeAs plane, Fe$_2$As-type AFeAs (111-type, A=Li or Na),~\cite{111-1,111-2} as well as anti-PbO-type Fe(Se,Te) (11-type).\cite{11-1,11-2} The pairing mechanism and order parameter symmetry in new superconductors became important issues. The $s_{\pm}$ superconductivity was proposed in which the sign of the order parameter is switched between the two sets of Fermi surfaces.\cite{pairing1,pairing2} The multi-band electronic structure and the inter-pocket hopping or Fermi surface nesting seem to be common ingredients in all iron-based superconductors. \cite{pairing1,pairing2,pairing3} However, after more iron-based superconductors have been observed, the consensus is further away.\cite{pairing4,pairing5,pairing6,pairing7}

Recently, a new series of iron-based superconductors A$_x$Fe$_2$Se$_2$ (A=K,Rb,Cs,Tl) has been discovered with relatively high $T_c\sim 30$ K. \cite{kfese1,kfese2,kfese3,kfese4,kfese5} These compounds are purely electron-doped and it is found that the superconductivity might be in proximity of a Mott insulator.\cite{kfese3} The low-energy band structure and Fermi surface of K$_x$Fe$_{2-y}$Se$_2$ may be different from other iron-based superconductors. Only electron pockets were observed in angle-resolved photoemission experiment without a hole Fermi surface near the zone center. \cite{arpes1,arpes2,arpes3} This indicates that the sign change and  $s_{\pm}$-wave pairing is not a fundamental property of iron-based superconductors.

Thermal transport measurement is an effective method to probe the superconducting state and transport properties in the normal state in high-$T_c$ cuprates
and iron-based superconductors.\cite{sefat,TEP-1,TEP-2,TEP-3,TEP-4,TEP-9,thermalconduct2}  In SmFeAsO$_{0.85}$,
the magnitude of TEP develops a broad peak above $T_c$ coupled with a metallic resistivity behavior. This is attributed to resonant phonon scattering between electron and hole pocket indicating a significant Fermi surface nesting in this system.\cite{TEP-1} TEP in doped BaFe$_2$As$_2$ also pointed to a significant modification of the Fermi surface at small electron doping stabilizing low-temperature
superconductivity. \cite{TEP-2} In Fe$_{1+y}$Te$_{1-x}$Se$_x$ system, TEP and Nernst coefficients provide evidence of the low-density and strongly correlated superconductor. \cite{TEP-3}
Finally, thermal conductivity at very low temperature provides insight into the superconducting gap structure. \cite{pairing4,thermalconduct2}

Here we report temperature and magnetic field dependent thermal transport of K$_{x}$Fe$_{2-y}$Se$_2$ single crystal. Peak in thermal conductivity ($\kappa(T)$) is observed
at about $\frac{T_c}{2}$ which is attributed to phonons. The thermoelectric power above $T_c$ exhibits nearly linear behavior and could be described well by the carrier diffusion mechanism in a wide temperature range. The Fermi temperature $T_F$ deduced from these measurements yields a smaller $T_c/T_F$ than the value in Fe$_{1+y}$Te$_{1-x}$Se$_x$ system and other
well known correlated superconductors implying weaker electronic correlation.

\section{Experimental\label{Experimental}}
K$_{0.65(3)}$Fe$_{1.41(4)}$Se$_{2.00(4)}$ (K-122) single crystals were synthesized by self-flux method as described elsewhere in detail.~\cite{hechang} Thermal and electrical transport measurement were conducted in Quantum Design PPMS-9. The sample was cleaved to a rectangular shape with dimension 5$\times$2 mm$^{2}$ in the \textit{ab}-plane and 0.3 $\mu$m thickness along the \textit{c}-axis. Thermoelectric power and thermal conductivity were measured using steady state method and one-heater-two-thermometer setup with silver paint contact directly on the sample surface. The heat and electrical current were transported within the \textit{ab}-plane of the crystal oriented by Laue camera, with magnetic field along the \textit{c}-axis and perpendicular to the heat/electrical current. The sample is very sensitive to oxygen in air, and air exposure exceeding 1 hour will result in significant surface oxidization seen by incomplete transition in $\rho(T)$ and $\rho> 0$ below $T_c$. The exposure to air of samples we measured was less than 20 minutes. Clear and complete superconducting transition seen in $\rho(T)$ and $S(T)$ confirmed that the surface of our samples was not oxidized. The relative error in our measurement for both $\kappa$ and $S$ was below $5\%$ based on Ni standard measured under identical conditions.

\section{Results and Discussions}

\begin{figure}
\includegraphics[scale=0.9] {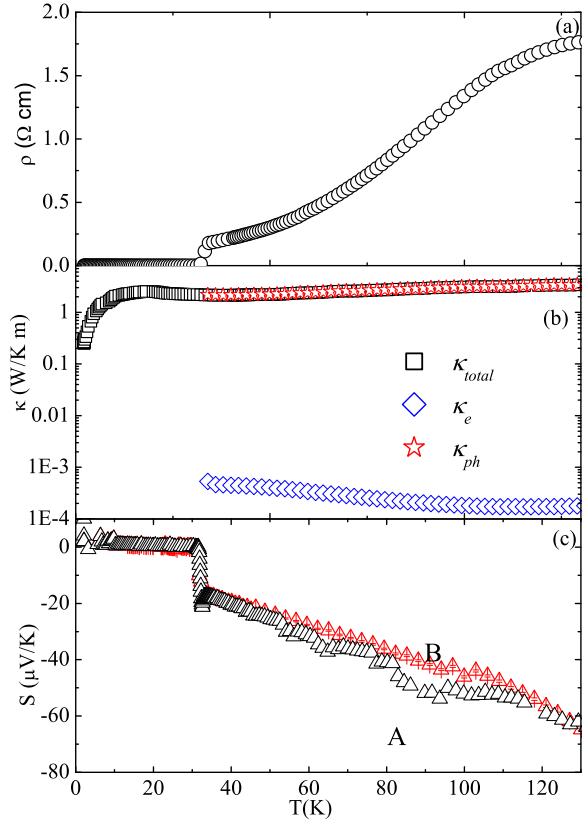}
\caption{(Color online) Temperature dependence of the resistivity (a), thermal conductivity (b) and thermoelectric power (c) for K-122 under zero magnetic field within temperature range from 2~K to 130~K. The electron term $\kappa_e$ is estimated using Wiedemann-Franz law and the phonon term $\kappa_{ph}$ is obtained by subtracting the electron term from total thermal conductivity (see text).  We show thermoelectric power $S(T)$ for two independently grown samples A and B.}
\end{figure}

Temperature dependence of the electrical resistivity $\rho(T)$, thermal conductivity $\kappa(T)$ and TEP $S(T)$ for K-122 in zero magnetic field between 2 K and 130 K, is shown in Fig. 1. The $\rho(T)$ is metallic below
$\sim125$ K, and superconducting below $\sim32$ K. These values are similar to previous report.\cite{kfese1,kfese2} Thermal
conductivity decreases with decrease in temperature in general showing a peak below $T_c$.  Thermoelectric power is negative, consistent with negative charge carriers. TEP results for two independently grown samples are nearly identical. The value of TEP decreases with decrease of temperature and exhibits nearly linear behavior up to 130 K. It vanishes at $T_c$ since Cooper pairs carry no entropy. The $T_c$ inferred from $S(T)=0$ for two samples is identical and is $31.8$ K, consistent with resistivity measurement.

Fig. 2(a) shows the thermal conductivity $\kappa$ and electrical resistivity $\rho$ near $T_c$. Below $T_c$ $\kappa$ increases with decrease in temperature and peaks near $\frac{T_c}{2}$ ($\sim$17 K). The peak in thermal conductivity below $T_c$ was observed in the hole-doped Ba$_{1-x}$K$_x$Fe$_2$As$_2$,\cite{TEP-5} electron-doped Ba(Fe$_{1-x}$Co$_x$)$_2$As$_2$,\cite{TEP-6} and other unconventional superconductors such as YBa$_2$Cu$_3$O$_{7-\delta}$\cite{TEP-7} and CeCoIn$_5$,\cite{TEP-8} where it was attributed to a large quasi-particle (QP) population and enhanced zero-field QP mean-free-path in the superconducting state. However, the resistivity in our sample is very high. Thermal conductivity is composed of the electron term $\kappa_e$ and the phonon term $\kappa_{ph}$; $\kappa_{total}=\kappa_e+\kappa_{ph}$. The electron term $\kappa_e$ above $T_c$, estimated using Wiedemann-Franz law $\frac{\kappa_e}{T}=\frac{L_0}{\rho}$, is very small and indicates a predominantly phonon contribution (Fig. 1(b)). In order to clarify the origin of this peak,  we show (Fig. 2(b)) the thermal conductivity of K$_{x}$Fe$_{2-y}$S$_2$ which is a semiconductor with identical crystal structure and somewhat reduced lattice parameters.\cite{kfes} K$_{x}$Fe$_{2-y}$S$_2$ also exhibits a peak in thermal conductivity at $\sim 22$ K that is obviously unrelated to $T_c$. Moreover, 9 T magnetic field suppresses superconductivity in K-122 to 27 K. \cite{hechang,apl} This should suppress the peak in thermal conductivity induced by QP, as seen in cuprate and Ba$_{1-x}$K$_x$Fe$_2$As$_2$.\cite{TEP-6,TEP-7} However, 9 T magnetic field has no significant influence on the peak observed in our sample, as shown by open circles in Fig. 2(a). The peak in $\kappa$ therefore is more likely to originate from phonons rather than the QP contribution. The phonon peak in lattice thermal conductivity is commonly found in materials due to the competition between the point-defect/boundary scattering and Umklapp phonon scattering mechanism.\cite{phononpeak}


\begin{figure}[tpb]
\centering
\includegraphics[scale=0.62]{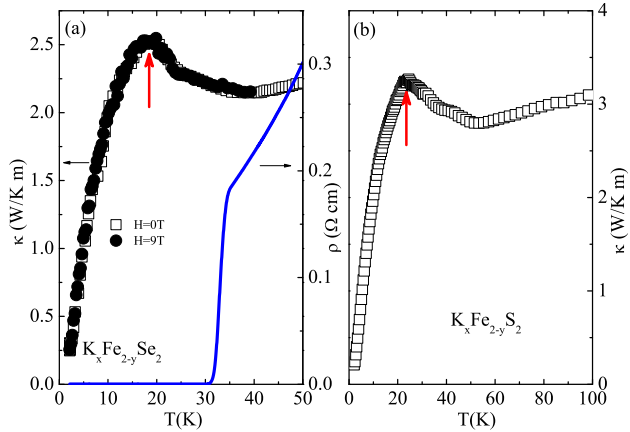}
\caption{(Color online) (a) Thermal conductivity in 0 T (open square) and 9 T (open circle) field, as well as electrical resisitivity in 0 T field (red line) of  K-122 near $T_c$. (b) Thermal conductivity in zero magnetic field for K$_{x}$Fe$_{2-y}$S$_2$. Red arrows point the position of the peak in thermal conductivity.}
\end{figure}

Fig. 3 present the  $\rho(T)$ and $S(T)$ near $T_c$ in different magnetic fields up to 9 T. The resistive transition strongly  broadens, and both the onset temperature and the zero-resistance are suppressed. Superconducting transition seen by TEP is also suppressed and broadened by external field but the amplitude of the normal state Seebeck response does not change for $\mu_0H <$ 9 T. These features are reminiscent of thermally induced motion of vortex lattices in superconductors. A small peak just above $T_c$ in TEP is observed only under zero field, as shown by the red arrow in Fig. 3(b). This is similar to some TEP observed in cuprates and is attributed to the AC measuring technique.\cite{cuprate1,ppms} The AC measurement technique picks up a voltage contribution of thermoelectric power derivative in addition to the linear term. This is present even for good thermal contacts and is reduced for lower density of data points and sharp transitions.

\begin{figure}
\includegraphics [scale=0.6]{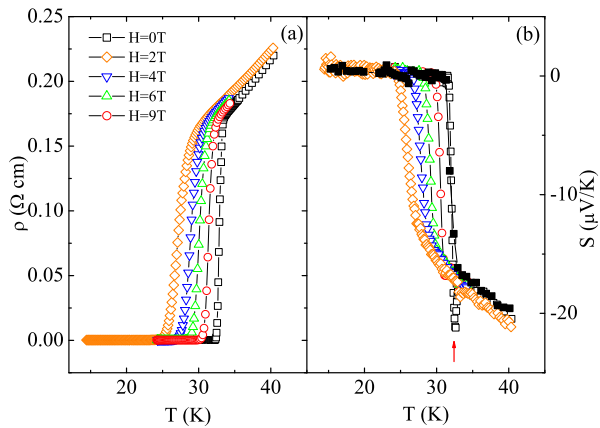}
\caption{(Color online) Temperature dependence of resistivity (a) and Seebeck coefficient (b) of K-122 in different magnetic field in temperature range between 2 K and 40 K. For TEP, zero field data of Sample B (close squares) is shown for comparison. The red arrow indicates the small peak position just above $T_c$ in TEP.}
\end{figure}

In the lower part of the transition ($\frac{\rho}{\rho_n}\leq1\%$), $\rho(T)$ is thermally activated according to Arrhenius law $\rho=\rho_0 exp\left(-\frac{U_a(B)}{k_BT}\right)$, where $U_a(B)$ is the activation energy for the flux motion.  A linear behavior occurs over a typical temperature interval of about 6K. The Seebeck coefficient follows identical temperature dependence in the same region. Fig. 4(a,b) show the Arrhenius plot $\ln(\rho)$ and $\ln(S)$ as a function of the inverse temperature. The conventional theory can not reproduce the large TEP in the mixed state, similar to the high-$T_c$ cuprates. \cite{cuprate11,cuprate2} The activation energy for the flux motion obtained from resistivity decreases exponentially with magnetic field, but the value obtained from TEP is nearly unchanged and larger, as shown in Fig. 4(c), also in contrast to conventional theory of flux motion.

\begin{figure}
\includegraphics[scale=0.43] {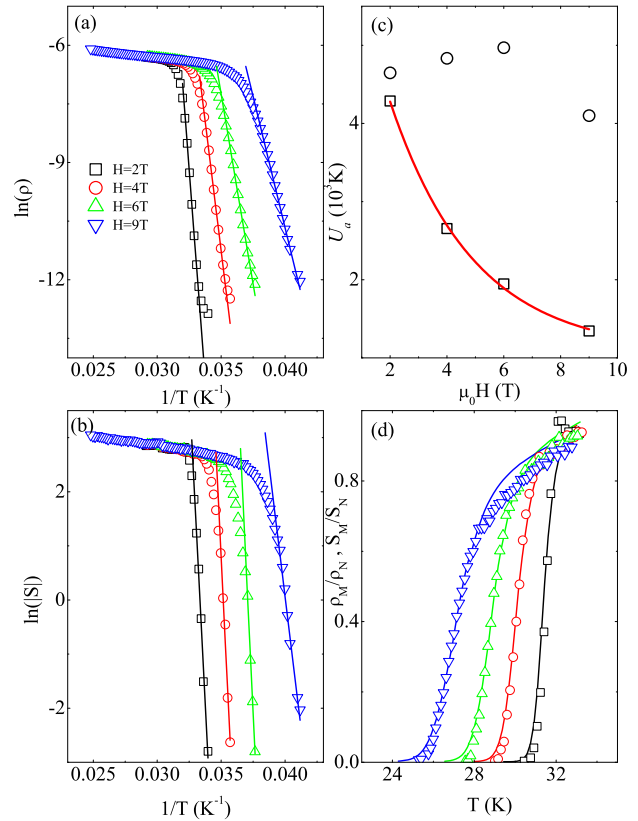}
\caption{(Color online) Arrhenius plot of resistivity (a) and Seebeck Coefficient (b). (c) The activation energy for flux motion derived from resistivity data (open squares) and TEP data (open circles) under different magnetic field. The red line is the exponential fitting results for the data from resistivity. (d) The relationship between $\frac{S_M}{S_N}$ (open symbols), $\frac{\rho_M}{\rho_N}$ (solid lines) and temperature under different magnetic field.}
\end{figure}

TEP is associated with entropy and heat transport parallel to the "induced" electric field. This longitudinal entropy flow is not carried by the normal excitations in the vortex cores since these excitations move with the vortices perpendicular to the induced electric field (Nernst effect). \cite{cuprate2} The longitudinal entropy transport has to be attributed to other excitations. We postulate that TEP in the mixed state near $T_c$ is attributed to quasi-particles (QP) excited over the energy gap as in high $T_c$ cuprates.\cite{cuprate11,cuprate2} Such excitations, not bound to the vortex core, are present in any superconductor at finite temperature and may be important in determining the dissipation due to the short coherence length.\cite{cuprate11,cuprate2} Therefore, an estimate of TEP can be obtained from
\begin{eqnarray}
\frac{S_M}{S_N}\simeq \frac{\rho_M}{\rho_N},
\end{eqnarray}
where $S_M,\rho_M,S_N,\rho_N$ are TEP and resistivity in the mixed state (M) and normal state (N), respectively, since TEP from longitudinal entropy flow has a small dependence on the Hall angle. \cite{cuprate2} This is in a good agreement with experimental data, as shown in Fig. 4(d).

We now turn to TEP in the normal state.  TEP is the sum of three different contributions: the diffusion term $S_{diff}$, the spin-dependent scattering term and the phonon-drag term $S_{drag}$ due to electron-phonon coupling.\cite{TEP-text,TEP-1} The spin dependent scattering or corresponding magnon drag effect always gives $\sim T^{\frac{3}{2}}$ dependence,\cite{magnon} which is not observed in our TEP results.  Moreover, TEP in our sample above $T_c$ is independent of magnetic field, which excludes the spin-dependent mechanism. The contribution of phonon-drag term often gives $\sim T^3$ dependence for $T<<\Theta_D$, $\sim 1/T$ for $T\geq\Theta_D$ (where $\Theta_D$ is the Debye Temperature), and a peak structure for $\sim\frac{\Theta_D}{5}$.\cite{phonon-drag1,phonon-drag2}  The estimated Debye temperature of K-122 system is about 260 K.\cite{hp} The absence of the peak structure in our TEP results suggests negligible contribution of the phonon drag effect to $S(T)$. Instead a nearly linear relationship is observed between 2 K and 120 K (Fig. 1(c)), suggesting that the diffusion term is dominant.

Diffusive Seebeck response of a Fermi liquid is expected to be linear in $T$ in the zero-temperature limit, with a magnitude proportional to the strength of electronic correlations.\cite{ratio} This is similar to the T-linear electronic specific heat, $C_e/T=\gamma$. Both can be linked to the Fermi temperature $T_F$:
\begin{eqnarray}
S/T & = & \pm\frac{\pi^{2}}{2}\frac{k_B}{e}\frac{1}{T_F} \\
\gamma & = & \frac{\pi^2}{2}k_B\frac{n}{T_F}
\end{eqnarray}
where $k_B$ is Boltzmann's constant, $e$ is the electron charge and $n$ is the carrier density.\cite{TEP-3} Fig. 5 presents the temperature dependence of TEP divided by $T$, $S/T$
under 0 T (open squares), 9 T (open circles) magnetic field for Sample A and 0 T (close squares) for Sample B, respectively. TEP in the normal state near $T_c$ is independent of magnetic field and can be described well by diffusive model. The zero-temperature extrapolated value of $S/T$
is $\sim0.48$ $\mu V/K$ for $\mu_0H<9$ T (Fig. 5). We can therefore extract $T_F=880$ K.

\begin{figure}
\includegraphics [scale=0.6]{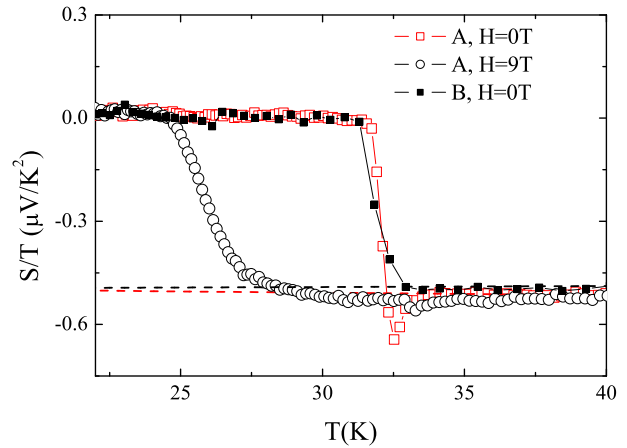}
\caption{(Color online) Temperature dependence of the Seebeck coefficient divided by $T$, $S/T$ in K$_{0.8}$Fe$_{2-y}$Se$_2$ under 0T (open squares) and 9T (open circles) magnetic field for sample A, and under 0T (close squares) for sample B, respectively. The dashed lines are linear fitting result within higher temperature range. \label{T-S/T}}
\end{figure}

\begin{table}[b]
\caption{Set of derived parameters for K$_{0.8}$Fe$_{2-y}$Se$_2$.}
\begin{ruledtabular}
\begin{tabular}{cc}
Quantity & Magnitude  \\
$k_F$ (nm$^{-1}$) & 2.6  \\
$\xi$ (nm) & 1.8 \\
$m^*$ ($m_e$) & 3.4 \\
$v_F$ (km/s) & 89
\end{tabular}
\end{ruledtabular}
\end{table}

The ratio of the superconducting transition temperature
to the normalized Fermi temperature $\frac{T_c}{T_F}$ characterizes the correlation strength in superconductors. In unconventional superconductors, such as CeCoIn$_5$\cite{CeCoIn} and YBa$_2$Cu$_3$O$_{6.67}$,\cite{YBCO} this ratio is about 0.1, but it is only $\sim0.02$ in BCS superconductors, such as LuNi$_2$B$_2$C. \cite{TEP-3} In Fe$_{1+y}$Te$_{1-x}$Se$_x$, $\frac{T_c}{T_F}$ is also near 0.1, pointing to importance of electronic correlations.\cite{TEP-3} When compared to these strongly correlated superconductors, $\frac{T_c}{T_F}\sim0.04$ in our crystal is relatively small, but is larger than in conventional superconductors. This implies that K-122 is a weakly or intermediately correlated superconductor. K-122 is proposed to be the first iron-based high temperature superconductor near an insulating antiferromagnetic order, just like cuprates whose parent compound is a Mott insulator and where the correlation effects dominate. \cite{kfese3} However, TEM study demonstrated the presence of ordered Fe vacancies in \textit{ab}-plane \cite{TEM} and theoretical study pointed out that the ordered Fe vacancies could induce the band narrowing and consequently decrease the correlating strength needed for Mott transition \cite{order}. This can explain the relatively weak correlation strength observed in our experiment.

Recent specific heat measurement \cite{hp} find $\gamma_n=6\pm0.5$ mJ/molK$^2$, which is smaller than in iron-based superconductors. The absolute value of the dimensionless ratio of TEP to specific heat $q=\frac{N_{Av}eS}{T\gamma}$, with $N_{Av}$ the Avogadro number, provides the carrier density.\cite{ratio} Calculation gives the carrier density with $|q|^{-1}\simeq 0.13$ carrier per unit cell, which is somewhat larger than the value in Fe$_{1+y}$Te$_{0.6}$Se$_{0.4}$.\cite{TEP-3} Given the volume of unit cell, we obtain the carrier density in volume $n\simeq 6.1\times10^{20}$ $cm^{-3}$ and derive Fermi momentum $k_F=(3\pi^2n)^{\frac{1}{3}}\simeq 2.6$ nm$^{-1}$. Ultimately we can derive effective mass $m^*$, Fermi velocity $v_F$, as well as superconducting coherence length $\xi$ using: $k_BT_F=\frac{\hbar^2k_F^2}{2m^*}, \hbar k_F=m^*v_F$, and $\xi=\frac{\hbar v_F}{\pi\Delta_0}$ with measured $\Delta_0=10.3$ meV by ARPES. \cite{arpes1} The results are listed in Table I. It is worth to note that $\xi$ can also be derived from the upper critical field $H_{c2}$, using $H_{c2}(0)=0.693[-\frac{dH_{c2}}{dT}]_{T_c}T_c$ and $\xi^{-2}=\frac{2\pi}{\Phi_0}\frac{H_{c2}(0)}{T_c}$. From Ref.\onlinecite{kfese2}, $[-\frac{dH_{c2}}{dT}]_{T_c}=3.17$ T/K, and the derived superconducting coherence length is $\xi\simeq 2.2$ nm consistent with value in Table I. This confirms the consistence of our derived parameters.

\section{Conclusion}
Thermal transport measurement of iron-based superconductor K$_x$Fe$_{2-y}$Se$_2$ have been performed on a single crystalline sample. The peak anomaly in thermal conductivity observed
at nearly $\frac{T_c}{2}\sim17$ K is attributed to phonon contribution. The large thermoelectric power in the mixed state could imply the large quasi-particle excitations over the energy gap. The thermoelectric power above $T_c$ exhibits nearly linear behavior and could be described well by the carrier diffusion mechanism in a wide temperature range. The zero-temperature extrapolated thermoelectric power
is smaller when compared to other correlated superconductors pointing to a large normalized Fermi temperature. These findings indicate that K$_x$Fe$_{2-y}$Se$_2$ is a weakly or intermediately correlated superconductor. The ordered Fe vacancies could induce the band narrowing and then decrease the correlated strength needed for Mott transition, as predicted by theory. We became aware that a preprint was posted on the arXiv.org with similar thermopower data in zero field on the same day of our submission. \cite{rongwei}

\begin{acknowledgments}
We acknowledge valuable discussions with Louis Taillefer. We thank John Warren for help with SEM measurements. Work at Brookhaven is supported by the U.S. DOE under contract No. DE-AC02-98CH10886 and in part by the center for Emergent Supercondcutivity, and Energy Frontier Research Center funded by the U.S. DOE, office for Basic Energy Science.
\end{acknowledgments}


\end{document}